\documentclass[preprint2,times,tighten]{aastex}

\usepackage{amsmath,amstext}

\newcommand{\iterations}{10,000 }
\newcommand{\lititerations}{10,000 }
\newcommand{\falsePosIterations}{1000 }
\newcommand{\falsePosIterationsTotal}{6000 }

\usepackage{ragged2e}
\usepackage{graphicx}
\usepackage{textcomp}
\usepackage{booktabs}
\usepackage{afterpage}
\usepackage{placeins}
\usepackage{amsmath}
\usepackage{multirow}
\usepackage[caption=false]{subfig}
\usepackage{tabularx}
\usepackage{scrextend}
\usepackage{hyperref}

\shorttitle{Variability of Sgr A* in X-ray and IR}
\shortauthors{Boyce et al.}

\begin{document}

\title{Simultaneous X-ray and Infrared Observations of Sagittarius A*'s Variability}
\author{H. Boyce\altaffilmark{1}, D. Haggard\altaffilmark{1,2}, G. Witzel\altaffilmark{3,4}, S. P. Willner\altaffilmark{5}, J. Neilsen\altaffilmark{6}, J. L. Hora\altaffilmark{5}, S. Markoff\altaffilmark{7}, G. Ponti\altaffilmark{8}, F. Baganoff\altaffilmark{9}, E. Becklin\altaffilmark{3}, G. Fazio\altaffilmark{5}, P. Lowrance\altaffilmark{10}, M. R. Morris\altaffilmark{3}, H. A. Smith\altaffilmark{5}}
\email{hope.boyce@mail.mcgill.ca}
\altaffiltext{1}{Department of Physics and McGill Space Institute, McGill University, 3600 University St., Montreal QC, H3A 2T8, Canada}
\altaffiltext{2}{CIFAR Azrieli Global Scholar, Gravity \& the Extreme Universe Program, Canadian Institute for Advanced Research, 661 University Ave., Suite 505, Toronto, Ontario M5G 1M1, Canada}
\altaffiltext{3}{University of California, Los Angeles, CA 90095, USA}
\altaffiltext{4}{Max Planck Institute for Radio Astronomy, Bonn, Germany}
\altaffiltext{5}{Harvard-Smithsonian Center for Astrophysics, 60 Garden St., MS-65, Cambridge, MA 02138, USA}
\altaffiltext{6}{Kavli Institute for Astrophysics \& Space Research, MIT, 70 Vassar St., Cambridge, MA 02139, USA}
\altaffiltext{7}{Anton Pannekoek Institute for Astronomy/GRAPPA, University of Amsterdam, 1098 XH Amsterdam, The Netherlands}
\altaffiltext{8}{Max-Planck-Institut f\"{u}r extraterrestrische Physik, Giessenbachstra{\ss}e, 85748 Garching, Germany}
\altaffiltext{9}{Massachusetts Institute of Technology, 77 Massachusetts Avenue, 37-555, Cambridge, MA 02139, USA}
\altaffiltext{10}{IPAC-Spitzer, MC 314-6, California Institute of Technology, Pasadena, CA 91125, USA}

\begin{abstract}
Emission from Sgr A* is highly variable at both X-ray and infrared (IR) wavelengths. Observations over the last $\sim$20 years have revealed X-ray flares that rise above a quiescent thermal background about once per day, while faint X-ray flares from Sgr A* are undetectable below the constant thermal emission. In contrast, the IR emission of Sgr A* is observed to be continuously variable. Recently, simultaneous observations have indicated a rise in IR flux density around the same time as every distinct X-ray flare, while the opposite is not always true (peaks in the IR emission may not be coincident with an X-ray flare). Characterizing the behaviour of these simultaneous X-ray/IR events and measuring any time lag between them can constrain models of Sgr A*'s accretion flow and the flare emission mechanism. Using 100+ hours of data from a coordinated campaign between the \textit{Spitzer Space Telescope} and the \textit{Chandra X-ray Observatory}, we present results of the longest simultaneous IR and X-ray observations of Sgr A* taken to date. The cross-correlation between the IR and X-ray light curves in this unprecedented dataset, which includes four modest X-ray/IR flares, indicates that flaring in the X-ray may lead the IR by approximately 10--20 minutes with 68\% confidence. However, the 99.7\% confidence interval on the time-lag also includes zero, i.e., the flaring remains statistically consistent with simultaneity. Long duration and simultaneous multiwavelength observations of additional bright flares will improve our ability to constrain the flare timing characteristics and emission mechanisms, and must be a priority for Galactic Center observing campaigns.
\end{abstract}

\maketitle

\section{Introduction}
Sagittarius A* (Sgr A*) sits at the center of the Milky Way and is the closest supermassive black hole (SMBH), with a distance of only $\sim$$8$ kpc \citep[e.g.,][]{2016ApJ...830...17B,2017ApJ...837...30G}. Monitored in the radio since its discovery, and more recently in the infrared (IR) and the X-ray, Sgr A* has a mass of $\sim$$4\times10^{6}$ M$_{\sun}$ \citep{2016ApJ...830...17B,2017ApJ...837...30G}, an extremely low bolometric-to-Eddington luminosity ratio \citep[$L/L_{Edd}\sim10^{-9}$;][]{2010RvMP...82.3121G} and appears to be accreting material at a very low rate \citep[$\lesssim$$10^{-7}$M$_{\sun}$ yr$^{-1}$;][]{2003ApJ...591..891B,2006ApJ...640..308M,2007ApJ...654L..57M,2012ApJ...755..133S, 2015ApJ...809...10Y}. 

\par In X-rays, Sgr A* appears as a persistent source, with a flux of about 3$\times$10$^{33}$ erg/s \citep{2001Natur.413...45B,2003ApJ...591..891B}. This faint, steady emission arises from thermal Bremsstrahlung radiation from a hot accretion flow dominated by regions near the Bondi radius \citep{2002ApJ...575..855Q, 2003ApJ...591..891B, 2003ApJ...598..301Y, 2004ApJ...611L.101L, 2006ApJ...640..319X, 2013Sci...341..981W} and is interrupted about once per day by distinct flares of non-thermal emission coming from very close to the black hole \citep{2013ApJ...774...42N, 2015ApJ...799..199N}. First detections of Sgr A* in the IR also revealed a highly variable source \citep{2003Natur.425..934G, 2004ApJ...601L.159G} with peaks in the IR emission detected more frequently than in X-rays. Since these first discoveries, the statistical behaviour of both the X-ray \citep[e.g.,][]{2015ApJ...799..199N,2015ApJ...810...19L,2015MNRAS.454.1525P} and the IR \citep[e.g.,][]{2011ApJ...728...37D,2014ApJ...793..120H, 2012ApJS..203...18W, 2018ApJ...863...15W} activity have been well studied and the flux density distributions of these two wavelengths can both be described by a power law \citep{2015ApJ...799..199N,2012ApJS..203...18W, 2018ApJ...863...15W}. Though the X-ray and IR have no other clear statistical similarities, the coincidence of peaks in the variability hint that there may be a physical connection between them \citep[e.g.,][]{2006A&A...450..535E,2006ApJ...644..198Y,2008A&A...479..625E,2009ApJ...698..676D,2009ApJ...706..348Y,2016A&A...589A.116M}. The picture is even less clear at longer wavelengths, where some submm flares have been tentatively linked to IR flares with some delay \citep{2008ApJ...682..373M,2012RAA....12..995M} or no delay \citep[][]{2018ApJ...864...58F}, and radio variability has not been observed to coincide with IR or X-ray activity \citep[e.g.,][]{2017ApJ...845...35C}. However, upcoming submm and coordinated multi-wavelength observations undertaken by the Event Horizon Telescope collaboration \citep{2008Natur.455...78D} may shed light on these connections in the near future.

\par Despite the intensive efforts that have been made to characterize Sgr A*'s variable emission, the physical mechanisms producing the variability are still unknown. Suggested physical models include particle acceleration due to magnetic reconnection events, violent disk instabilities, jets, other stochastic processes in the accretion flow \citep[e.g.,][]{2001A&A...379L..13M, 2002ApJ...566L..77L, 2003ApJ...598..301Y, 2004ApJ...611L.101L,2009ApJ...703L.142D,2009A&A...508L..13M,2010ApJ...725..450D,2016ApJ...826...77B,2017MNRAS.468.2552L}, and even tidal disruption of asteroids \citep{2008A&A...487..527C, 2009A&A...496..307K, 2012MNRAS.421.1315Z}. Additional models attempt to explain the variability in the context of expanding plasma blobs \citep[e.g.,][]{1966Natur.211.1131V, 2008ApJ...682..373M, 2015MNRAS.454.3283Y, 2017MNRAS.468.2552L}, themselves launched by magnetic reconnection events or unsteady jet emission.  Finally, gravitational lensing near the horizon of the SMBH might add an amplifying effect to the observed emission \citep{2015ApJ...812..103C}. 

\par One possible picture for the IR flares describes a population of electrons undergoing continuous acceleration due to turbulent processes in the inner accretion flow and subsequently emitting synchrotron radiation. This is supported by observed timescales for the IR variability, with factors of $\gtrsim$10 changes within $\sim$10 minutes \citep[e.g.,][]{2003Natur.425..934G, 2004ApJ...601L.159G,2018ApJ...863...15W}, the spectral index at high flux densities \citep[$\alpha \approx 0.6$;][]{2007ApJ...667..900H, 2011A&A...532A..26B, 2014IAUS..303..274W}, and the high linear polarization of the IR emission \citep{2006A&A...455....1E, 2006A&A...460...15M, 2007A&A...473..707M, 2007MNRAS.375..764T, 2007ApJ...668L..47Y, 2008A&A...479..625E, 2011A&A...525A.130W, 2015A&A...576A..20S}. The exact physical parameters of this turbulant acceleration of electrons (e.g., background magnetic field strength and $\gamma$, the Lorentz factor of the electrons) and the details of the radiative processes linking the IR flares to the X-ray remain unknown.

\par The radiation mechanisms often invoked to connect IR flares to simultaneous X-ray outbursts include pure synchrotron \citep[e.g.,][]{2001A&A...379L..13M,2009ApJ...698..676D, 2014ApJ...786...46B, 2017MNRAS.468.2447P}, synchrotron self-Compton \citep{2001A&A...379L..13M,2008A&A...479..625E}, and inverse Compton \citep{2012AJ....144....1Y}. Monitoring the black hole \textit{simultaneously} in multiple wavelengths can constrain these variability models via the association of peaks at different energies and times. With this motivation, Sgr A* has recently been monitored simultaneously at wavelengths ranging from the radio, mm, submm, infrared (IR), X-rays, to gamma-rays \citep[e.g.,][]{2009ApJ...706..348Y,2011A&A...528A.140T, 2012A&A...537A..52E, 2006A&A...450..535E, 2006ApJ...644..198Y,2009ApJ...698..676D,2018ApJ...864...58F}. 

\par Sgr A*'s flaring radiation mechanisms can also be constrained by the statistical behaviour of the flares. For example, \cite{2016MNRAS.461..552D} showed that the measured flux distributions of flares in the X-ray \citep{2013ApJ...774...42N, 2015ApJ...799..199N} and IR \citep{2012ApJS..203...18W, 2018ApJ...863...15W} were consistent with both synchrotron and SSC models, though their different shapes are difficult to understand if driven by the same particle populations.

The first simultaneous IR and X-ray observations of Sgr A* were carried out by \cite{2004A&A...427....1E}. Since then, studies with both X-ray and IR monitoring of Sgr A* have often reported that flares in the two wavelengths are simultaneous with each other or with the X-ray leading the IR by no more than 10 minutes \citep{2006A&A...450..535E,2006ApJ...644..198Y,2008A&A...479..625E,2009ApJ...698..676D,2009ApJ...706..348Y,2016A&A...589A.116M}. The exception is \cite{2012AJ....144....1Y}, who reported a short time lag between the maxima of the infrared and X-ray flares, with the X-ray flares possibly \textit{lagging} the IR. Such a behaviour would point toward an interpretation of an inverse Compton scattering model, where a fraction of IR photons are up-scattered to X-ray energies and are seen as an X-ray ``echo". 

\par These studies support the inference of a physical connection between the X-ray and IR flaring. However, the ground-based IR observations are often significantly shorter than the X-ray observations and there are frequently gaps in the data. When flares occur in one wavelength outside the observing window of another observatory, it is difficult to robustly associate two events \citep[e.g.,][]{2008ApJ...682..373M, 2016A&A...589A.116M,2017ApJ...845...35C}, leading to uncertainty in the cross-correlation. 

\par The first observations of Sgr A* with the \textit{Spitzer Space Telescope} \citep{2014ApJ...793..120H} provided a continuous $>$23 hour light curve of Sgr A* at infrared wavelengths that was more than a factor of two longer than the previous record holder \citep[600 minutes;][]{2008ApJ...688L..17M}. Building on this study, we utilized two space telescopes (\textit{Spitzer} and \textit{Chandra}) to obtain six simultaneous observations of Sgr A* at 4.5$ \mu$m and 2--8 keV. Four of these epochs have $\sim$24 hours of overlapping coverage from the two observatories, maximizing the probability of catching the relatively rare X-ray flares with simultaneous IR monitoring. A detailed statistical analysis of the IR data was presented in \cite{2018ApJ...863...15W}, and our first multi-wavelength results were reported in \cite{2018ApJ...864...58F}. 

\par In this work we investigated the temporal correlations between X-ray and IR variability using these six contemporaneous \textit{Chandra} and \textit{Spitzer} observations of Sgr A*. Section \ref{Observations} describes these observations and the reduction of the data, while Section \ref{Analysis} details our characterization of the variability of Sgr A* by cross-correlating the light curves. Section \ref{Discussion} explores the results in the context of previous studies and discusses their implications for models of the variability.

\begin{table*}[t]
  \begin{center}
    \caption{Observing log for simultaneous \textit{Chandra} and \textit{Spitzer} observations.}
    \label{tab:Observations}
    \begin{tabular}{l*{7}{c}} 
      \toprule
       & \multicolumn{3}{c}{\textit{Chandra}}  & \multicolumn{4}{c}{\textit{Spitzer}} \\
       \cmidrule(lr){2-4}
       \cmidrule(lr){5-8}
       Obs Date & ObsID & Obs. Start (UT)\footnote{\label{plus1}A (+1) in the start/end time columns indicates the offest of one day from the first date listed in the first column. For example, the first Chandra observation began at 02:59:23 on June 3rd. Times are UTC at the observatory. The time-lag analysis was based on corrected heliocentric times.} & Obs. End (UT)\footref{plus1} & AORKEY & AOR Start (UT)\footref{plus1} & AOR End (UT)\footref{plus1} & Mode\footnote{Mode of operation (either Map or Stare modes). The ``Map" mode was a short operation performed after the initial slew to Sgr A*, while the two ``Stare" modes were each $\sim$12 hour long observations taken with a $\sim$4 min break between the two.} \\
       \midrule
       \multirow{3}{*}{2014 June 2} & \multirow{3}{*}{16210} & \multirow{3}{*}{(+1) 02:59:23} & \multirow{3}{*}{(+1) 08:40:34} & 51040768 & 22:32:00 & 22:56:01 & Map \\
       & & & & 51041024 & 22:59:37 & (+1) 10:39:44 & Stare part 1 \\
       & & & & 51041280 & (+1) 10:43:22 & (+1) 22:23:28 & Stare part 2 \\
       \midrule
       \multirow{3}{*}{2014 July 4} &  \multirow{3}{*}{16597} & \multirow{3}{*}{20:48:12} & \multirow{3}{*}{(+1) 02:21:32} & 51344128 & 13:21:59 & 13:45:55 & Map \\
       & & & & 51344384 & 13:49:37 & (+1) 01:29:43 & Stare part 1 \\
       & & & & 51344640 & (+1) 01:33:21 & (+1) 03:13:27 & Stare part 2 \\
       \midrule
       \multirow{3}{*}{2016 July 12} &  \multirow{3}{*}{18731} & \multirow{3}{*}{18:23:59} & \multirow{3}{*}{(+1) 18:42:51} & 58115840 & 18:04:23 & 18:34:03 & Map \\
       & & & & 58116352 & 18:37:45 & (+1) 06:37:30 & Stare part 1 \\
       & & & & 58116608 & (+1) 06:41:14 & (+1) 18:40:58 & Stare part 2 \\
       \midrule
       \multirow{3}{*}{2016 July 18} &  \multirow{3}{*}{18732} & \multirow{3}{*}{12:01:38} & \multirow{3}{*}{(+1) 12:09:00} & 58116096 & 11:44:02 & 12:13:43 & Map \\
       & & & & 58116864 & 12:17:25 & (+1) 00:17:09 & Stare part 1 \\
       & & & & 58117120 & (+1) 00:20:54 & (+1) 12:20:38 & Stare part 2 \\
       \midrule
       \multirow{3}{*}{2017 July 15} &  \multirow{3}{*}{19703} & \multirow{3}{*}{22:36:07} & \multirow{3}{*}{(+2) 00:01:34} & 60651008 & 22:28:54 & 22:58:34 & Map \\
       & & & & 63303680 & 23:02:17 & (+1) 11:02:01 & Stare part 1 \\
       & & & & 63303936 & (+1) 11:05:46 & (+1) 23:05:30 & Stare part 2 \\
       \midrule
       \multirow{3}{*}{2017 July 25} &  \multirow{3}{*}{19704} & \multirow{3}{*}{22:57:27} & \multirow{3}{*}{(+1) 23:28:30} & 60651264 & 22:39:33 & 23:09:14 & Map \\
       & & & & 63304192 & 23:12:57 & (+1) 11:12:41 & Stare part 1 \\
       & & & & 63304448 & (+1) 11:16:26 & (+1) 23:16:10 & Stare part 2 \\
       \bottomrule
    \end{tabular}
  \end{center}
\end{table*}

\vspace{8mm}
\section{Observations and Data Reduction} \label{Observations}
The IRAC instrument \citep{2004ApJS..154...10F} on the \textit{Spitzer Space Telescope} \citep{2004ApJS..154....1W} was used to observe Sgr A* at 4.5 $\mu$m for eight $\sim$24-hour long stretches between 2013 and 2017. Six of these observations had simultaneous monitoring from the \textit{Chandra X-ray Observatory} \citep{2000SPIE.4012....2W}, centered on Sgr A*'s radio position \citep[RA, Dec $=$ 17:45:40.0409, $-$29:00:28.118;][]{2004ApJ...616..872R} and are listed in Table \ref{tab:Observations}. The first two epochs had overlapping \textit{Chandra} coverage for $\sim$5.5 hours, while the other four had continuous  $\sim$24 hour coverage from both observatories. This results in a total of $\sim$107 hours for which we collected simultaneous X-ray and IR data -- the largest such data set to date.  Figure \ref{fig:lightcurves} shows the resulting Sgr A* light curves.

\subsection{Spitzer}
Sgr A* and its immediate surroundings are unresolved with the \textit{Spitzer}/IRAC detector. The measured flux of a single pixel is also highly sensitive to any changes in the telescope's pointing, even on the subpixel level. This makes detecting the intrinsic variability of Sgr A* a significant challenge. Recovering the signal requires modelling the flux variations of the pixel containing the black hole as a property of both 1) the varying intra-pixel sensitivity of \textit{Spitzer}/IRAC detector and 2) the telescope's sub-pixel motion. \cite{2014ApJ...793..120H} presented the first detection of Sgr A* with \textit{Spitzer} and demonstrated that a very precise relative flux measurement can be recovered for Sgr A* using this approach.

\begin{figure*}
\plotone{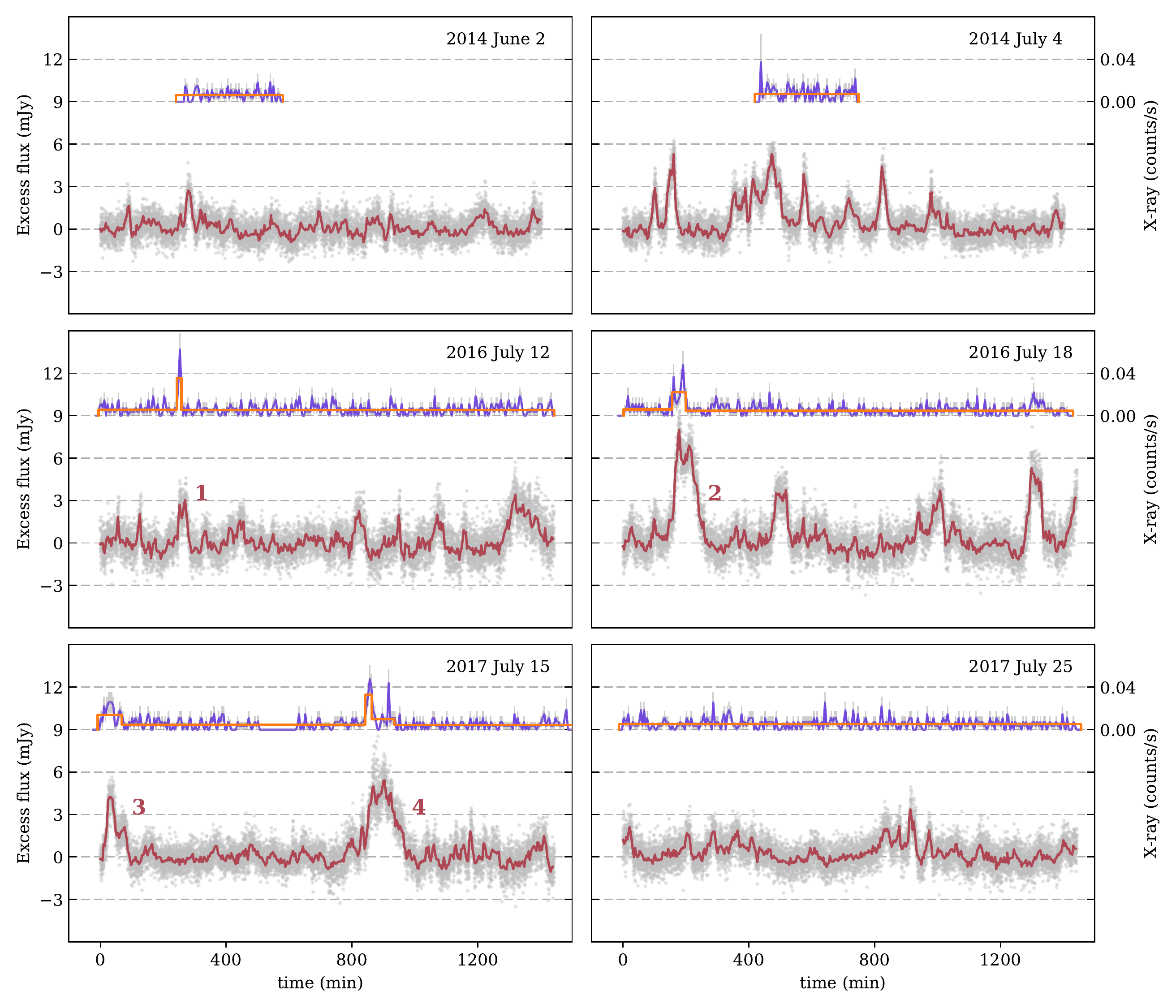}
\caption{Simultaneous IR and X-ray light curves of Sgr A*. Plotted in grey/red is the excess flux density (mJy) of the pixel containing Sgr A* from \textit{Spitzer} 4.5$\mu$m observations (see 2.1 of \cite{2018ApJ...863...15W}). Times for each epoch are relative to the beginning of the \textit{Spitzer} observations, measured in Heliocentric Modified Julian Date. Grey dots are the flux densities of each 6.4 s BCD coadd, while the red line shows the data binned over 3.5 minutes. \textit{Chandra} light curves of Sgr A* at 2--8 keV are plotted in purple with 300 s binning. The p$_{0}=0.05$ Bayesian Blocks results are overplotted on the X-ray curves in orange. Labels 1--4 indicate the four IR flux peaks associated with significant X-ray activity.}
\label{fig:lightcurves}
\end{figure*}

\par The \textit{Spitzer} data presented in this work were obtained and reduced using the procedures described by \cite{2014ApJ...793..120H} and \cite{2018ApJ...863...15W}, the latter of which analyzes these and additional \textit{Spitzer} light curves in the context of other IR datasets from Keck and the VLT. As documented in Table \ref{tab:Observations}, each observation was conducted as a set of three Astronomical Observation Requests (AORs). Each epoch followed the same observing sequence: an initial mapping operation performed after the slew to the Sgr A* field followed by two successive staring operations. Each staring operation began by using the ``PCRS Peakup" mode to position Sgr A* on the center of pixel (16,16) in the IRAC subarray. The subarray mode for \textit{Spitzer}/IRAC reads out 64 consecutive images (a ``frame set") of a 32x32 pixel region on the IRAC detector. This frame set is known as one Basic Calibrated Data product (BCD), which is the data format downloaded from the \textit{Spitzer} Heritage Archive\footnote{The \textit{Spitzer} Heritage Archive (\url{http://irsa.ipac.caltech.edu}) is part of the NASA/ IPAC Infrared Science Archive, which is operated by the Jet Propulsion Laboratory, California Institute of Technology, under contract with the National Aeronautics and Space Administration.}. Each frame in the frame set is a 0.1 s 32$\times$32 image, so one frame set takes 6.4 s to complete. After converting the pixel intensity into mJy, each frame set was combined into a single 32$\times$32 image referred to as a ``6.4 s BCD coadd". Since frame sets were typically separated by 2 s of telescope overheads, this resulted in an observation cadence of approximately 8.4 s.

The data reduction is described in Section 2.1 of \cite{2018ApJ...863...15W}, which is an improved version of the procedure in Appendix A1 of \cite{2014ApJ...793..120H}. This procedure corrects for the effect of nearby sources on the measured flux of Sgr A* as the telescope jitters on a subpixel basis throughout the observations. The resulting light curves are displayed in Figure \ref{fig:lightcurves}. The baseline flux density of these IR light curves is unknown, though the value has been inferred to be 1.9 mJy from the cumulative distributions of flux densities of Sgr A* in \cite{2018ApJ...863...15W}. Therefore, the \textit{Spitzer} light curves in Figure \ref{fig:lightcurves} plot the excess flux density above 1.9 mJy from pixel (16,16), attributed to Sgr A*'s variability.

\vspace{5mm}
\subsection{Chandra}

All \textit{Chandra} observations were acquired using the ACIS-S3 chip in the FAINT mode with a 1/8 subarray. The small subarray was chosen to avoid photon pileup in any bright flares from Sgr A* and in the nearby magnetar, SGR J1745-2900 \citep{2015MNRAS.449.2685C, 2017MNRAS.471.1819C}. 
\par We performed \textit{Chandra} data reduction and analysis with CIAO v4.9 tools\footnote{Chandra Interactive Analysis of Observations (CIAO) software is available at http://cxc.harvard.edu/ciao/} \citep{2006SPIE.6270E..1VF} and calibration database 4.7.3. The \texttt{chandra\_repro} script was used to reprocess level 2 events files before the WCS coordinate system was updated (\texttt{wcs\_update}). We used the CIAO tool \texttt{axbary} to apply barycentric corrections to the event times. Finally, we extracted a 2--8 keV light curve from a circular region of radius 1.25'' centered on Sgr A*. The small extraction region and energy range isolate Sgr A*'s emission from the nearby magnetar \citep[e.g.,][]{2017MNRAS.471.1819C} and the diffuse X-ray background \citep[e.g.,][]{2003ApJ...591..891B, 2012ApJ...759...95N,2013Sci...341..981W}. These X-ray light curves are plotted in Figure \ref{fig:lightcurves}.

\section{Analysis} \label{Analysis}

\subsection{Flare Detection and Characteristics}\label{FlareAnalysis}
To detect and characterize X-ray flares, we used the Bayesian Blocks algorithm as described by \cite{1998ApJ...504..405S} and \cite{2013ApJ...764..167S} and provided by Peter K. G. Williams \citep[\texttt{bblocks};][]{2017ascl.soft04001W}. This algorithm takes an unbinned, filtered \textit{Chandra} events file as input and models the light curve as a sequence of blocks with constant rates. A single block characterizes a region of the light curve for which there is no significant variability, and a light curve with a flare is made up of multiple blocks of differing count rates separated by change points. The code adopts a geometric prior on the number of blocks, preventing over-fitting the light curve by favouring fewer blocks when fewer events are present in the \textit{Chandra} events file. We ran the algorithm requiring a detection significance of 95\% for a single change point (a false positive rate of p$_{0}=0.05$), which implies that the overall detection significance of a flare (at least two change points) is $1-$p$_{0}^{2}\simeq99.8\%$ \cite[see, e.g.][]{2012ApJ...759...95N,2013ApJ...774...42N}.
\par We detect four X-ray flares during the total overlapping coverage of X-ray and IR coverage, one on 2016 July 12, one on 2016 July 18, and two on 2017 July 15. Increasing p$_{0}$ to 0.1 resulted in the detection of five flares (where the narrow peak around 910 mins on 2017 July 15 is also identified as an individual flare). Both numbers are consistent with past measurements of the average number of X-ray flares from Sgr A* \citep[$\sim$1.1/day;][]{2015ApJ...799..199N,2015MNRAS.454.1525P}. The results of the Bayesian Blocks analysis are overplotted on the X-ray light curves in Figure \ref{fig:lightcurves}. The lowest block in each light curve characterizes the quiescent flux from Sgr A* and had an average count rate of 0.006 counts/s. All of the X-ray flares detected above this constant thermal emission are relatively faint ($\lesssim$45 total counts) and don't contain enough counts to extract spectral information. 
\par While the X-ray flares are detected as distinct peaks rising above a constant background \citep[that may be hiding fainter X-ray variability;][]{2013ApJ...774...42N}, the emission from Sgr A* at IR wavelengths is constantly varying. In the three epochs where we observe significant X-ray activity, the IR flux from Sgr A* rises above $\sim$2 mJy within tens of minutes of the peak in the X-ray (see top row of Figure \ref{fig:zdcf}). The IR activity associated with significant X-ray flares appears to rise at the same time as the X-ray but last for a longer time (FWHM$_{\rm IR}$ $\gtrsim$2 $\times$ FWHM$_{\rm X-ray}$). These longer IR ``flares" are labeled 1-4 in Figures \ref{fig:lightcurves} and \ref{fig:zdcf}, with flares 1-3 roughly associated with the first three X-ray flares and flare 4 associated with both the fourth and fifth X-ray flares identified with the Bayesian blocks analysis. There are also multiple IR peaks at $\sim$2 mJy where we see no significant X-ray emission (e.g., around minute 300 on 2014 June 2; minute 450 on 2014 July 4; minute 500 and 1300 on 2016 July 18; and minute 900 on 2017 July 25).

\vspace{5mm}

\subsection{Cross-Correlation}\label{CrossCorrelation}
To quantify lags between the peaks of potentially associated activity in the X-ray and IR, we used the z-transform discrete correlation function \citep[\texttt{ZDCF};][]{1997ASSL..218..163A}, a tool that estimates the cross-correlation function without penalty for having a sparse or unevenly sampled light curve. We used the {\sc Fortran 95} implementation\footnote{www.weizmann.ac.il/weizsites/tal/research/software/} of both the \texttt{ZDCF} and the maximum likelihood function used to estimate the location of the \texttt{ZDCF} peak described in \cite{2013arXiv1302.1508A}. 

\par The \texttt{ZDCF} is not sensitive to the relative amplitudes of the input light curves. This allows us to run the \texttt{ZDCF} without renormalizing the \textit{Chandra} and \textit{Spitzer} data. The \texttt{ZDCF} \textit{is} however, sensitive to the shape of the light curves, meaning that two flares with similar rise times, envelopes, and decay times will result in the ZDCF having a stronger correlation with less uncertainty in the time lag. In our case, this means that if a flare in the X-ray light curve has a significantly shorter duration than a coincident flare in the IR light curve, the peak in the \texttt{ZDCF} will be weaker and flatter that if they had the same duration, limiting the precision in the measured time lag. 

\par
\begin{figure*}[ht!]
\centering
\includegraphics[width=0.95\textwidth]{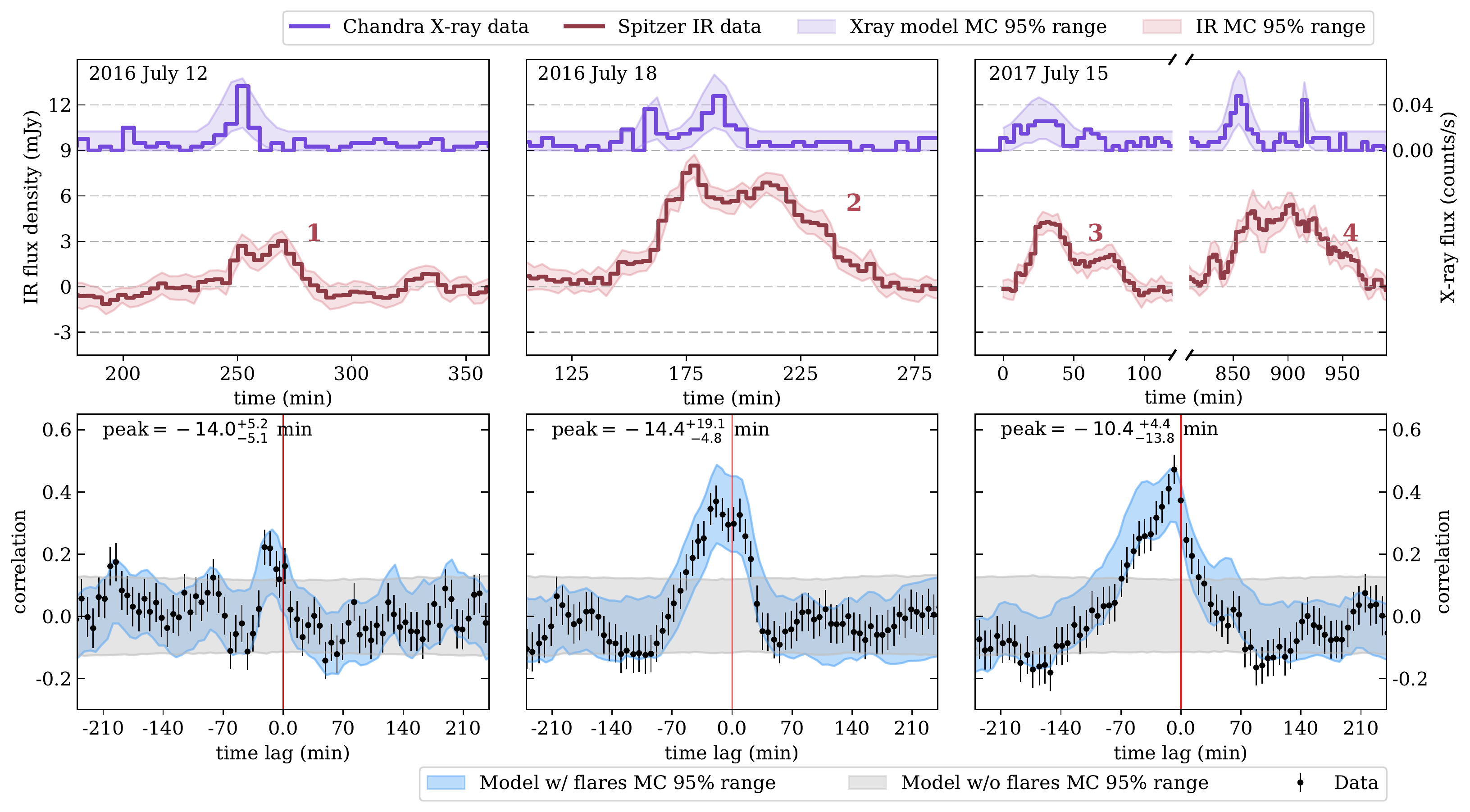}
\caption{Results from running the \texttt{ZDCF} on the three epochs that have X-ray flaring activity. The top row shows the data zoomed in on the portions of the light curves where we see significant X-ray activity. Labels 1--4 indicate the four IR peaks associated with this activity and also marked in Figure \ref{fig:lightcurves}. X-ray data are displayed in purple with 5 minute bins, and IR is displayed in red with 3.5 minute bins. Their respective envelopes show the 95\% range of the \iterations Monte Carlo (MC) realizations of the light curves. The third panel (2017 July 15) zooms in on two sections of the light-curve. The bottom row shows results of running the \texttt{ZDCF} on the entire $\sim$24 hour light curves for each of the dates in the top panels. The blue envelope is the 95\% range of results from running the MC realizations of the X-ray and IR light curves through the \texttt{ZDCF}, while the grey envelope is the 95\% range of the results of running the IR MC realizations through the \texttt{ZDCF} with \iterations realizations of simulated poisson noise consistent with the characteristics of the X-ray quiescent emission (no flares). The time lag and 68\% confidence interval from running \texttt{PLIKE} on the \iterations MC \texttt{ZDCF} results is displayed in the top left corner of these panels. The negative values for the position of the peaks indicate that the X-ray leads the IR.}
\label{fig:zdcf}
\end{figure*}

 For our analysis we used the 3.5 min binned IR light curves as the first input and the 300 s binned X-ray light curves as the second input. A positive time lag corresponds to a feature in the IR leading the X-ray, and a negative time lag corresponds to a feature in the IR lagging the X-ray. Figure \ref{fig:zdcf} shows the results of running the \texttt{ZDCF} on the three epochs for which we detect significant X-ray activity (2016 July 12, 2016 July 18, and 2017 July 15).
\par To identify significant correlation peaks, we generated \iterations Monte Carlo (MC) realizations of the X-ray and IR light curves and ran them through the \texttt{ZDCF}. These results are visualized as the blue envelope in the bottom row of Figure \ref{fig:zdcf}. The X-ray MC realizations were generated by adding Poisson noise to a smooth model containing the Gaussians fit to the flares. The IR realizations were generated from the 3.5-min-binned light curves by perturbing each point by a random amount drawn from a Gaussian distribution with standard deviation equal to the standard deviation of the mean of the $\sim$25 BCD coadds in the bin. We confirmed that these distributions in the Spitzer light curve bins are normal and dominated by white noise \citep[see][]{2018ApJ...863...15W}. For comparison, we cross-correlated the \iterations IR MC realizations with a separate set of \iterations X-ray MC light curves containing \textit{only} Poisson noise at the level of the quiescent flux (no flares, plotted in grey in Figure \ref{fig:zdcf}). Significant correlation peaks are those that rise above the grey MC ``no flare" envelope. There are correlation peaks near zero time lag in all three epochs. Peaks in the 2016 July 18 and 2017 July 15 epochs are considered significant and the small peak in the 2016 July 12 epoch is considered marginally significant. The highest points in the peak of these correlations all occur at negative time lags, implying that flares in the X-ray may \textit{lead} activity in the IR. 

\par To robustly measure the uncertainty on the time lags, we located the position of the peak in all \iterations results from \texttt{ZDCF} and defined confidence intervals based on the distribution of these \iterations time lags. This method for estimating the uncertainties takes the errors on the light curve data into account. Table \ref{tab:timeLags} compiles the time lags and the confidence intervals found in the MC analysis. All three epochs indicate the X-ray emission peaks approximately 10--20 min before the IR flux density peak.

\begin{table}[h!]
	\caption{Time Lags: \textit{Spitzer}/\textit{Chandra} Flares}
	\label{tab:timeLags}
\begin{tabularx}{0.475\textwidth}{@{\extracolsep{\fill}}lcrr}
	
	\toprule
	Date & time lag (min) & 68\% interval & 99.7\% interval \\
	\midrule
	2016 July 12 & $-14.0\substack{+5.2 \\ -5.1}$ & ($-19.1$,$-8.8$)  & ($-30.7$, $+2.8$) \\
	2016 July 18 & $-14.4\substack{+19.1 \\ -4.8}$ & ($-19.2$,$+4.7$) & ($-27.7$,$+19.2$) \\
	2017 July 15 & $-10.4\substack{+4.4 \\ -13.8}$ & ($-24.2$,$-6.0$) & ($-71.5$, $+6.4$) \\
	\bottomrule
	
\end{tabularx}

\textit{Note}: Negative values mean X-ray leads IR. Uncertainties on the time lag in the first column span the 68\% confidence interval on the \iterations MC runs. The second column displays the boundaries of this 68\% confidence interval, while the third column displays the 99.7\% confidence interval.
\end{table}

\par As an alternative analysis to the \texttt{ZDCF}, we also calculated the cross-power spectra of each epoch, but did not find it to show any strong trends or constrain any relevant timing between the two data sets. We also compared the \texttt{ZDCF} to the \texttt{ccf} function in R, which yielded almost identical results, but had the disadvantage of not requiring the two times series be sampled concurrently and at equally spaced points in time.

\section{Discussion} \label{Discussion}

\subsection{Are the X-ray and IR flares truly correlated?}

\par We detected IR activity nearly coincident with every X-ray flare, though not every IR peak with flux density level $\gtrsim$2 mJy had corresponding X-ray emission (perhaps because the weaker X-ray flares are hidden beneath the blanket of constant thermal emission). To investigate whether or not the observed X-ray and IR variabilities are truly associated, we consider the alternate possibility that the apparently correlated peaks in X-ray and IR emission are a chance association of peaks in typical X-ray and IR variability. 

\par We generated \falsePosIterations simulations of each of our six \textit{Spitzer} IR light curves by randomly drawing from the posterior of case 3 in \cite{2018ApJ...863...15W}, and then producing a random light curve for the given parameter set. The simulated light curves have the measurement noise properties of \textit{Spitzer}, and they are distributed accordingly to the log-normal flux density distribution determined for the M-band (4.5 $\mu$m). We ran the \falsePosIterations simulated light curves for each epoch through the \texttt{ZDCF} with the corresponding \textit{Chandra} X-ray light curves as the second input. This resulted in a total of \falsePosIterationsTotal cross-correlations between \textit{Chandra} X-ray light curves and simulated \textit{Spitzer} IR light curves. 
\par
\begin{figure}[h]
\plotone{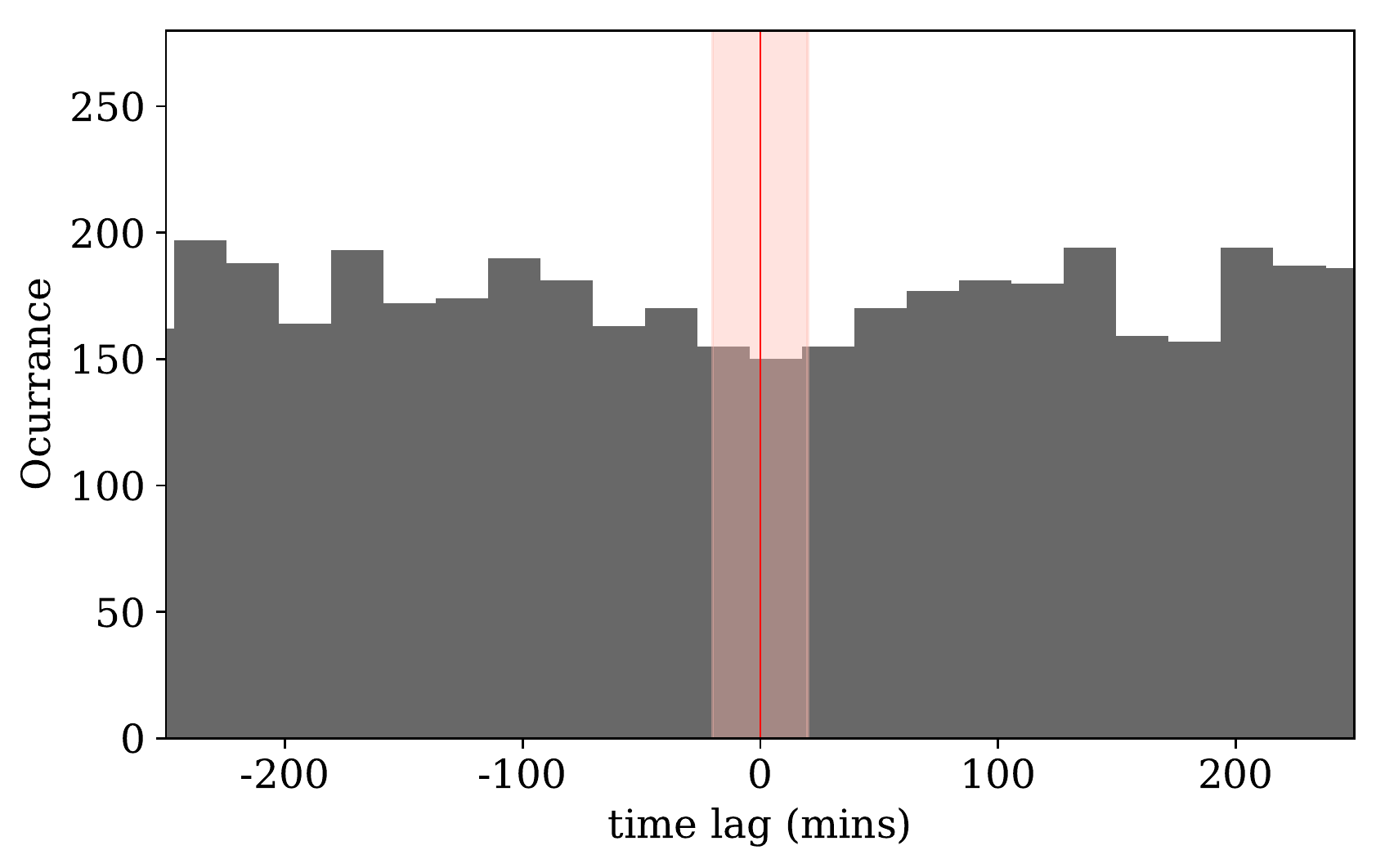}
\caption{Histogram of all the time lags measured in our \falsePosIterationsTotal simulations. The pink shaded region marks the range from -20 to +20 minutes and the thin red line marks zero time-lag.}
\label{fig:hist}
\end{figure}

 To test the probability of detecting a time lag similar to the one we measured by chance, we counted the occurrences of significant time lags measured between -20 and 0 minutes in our \falsePosIterationsTotal cross-correlations. A detection of a time lag within this window occurred in 138 of our \falsePosIterationsTotal instances. In other words, given \textit{Chandra} X-ray data and a random \textit{Spitzer} light curve simulating the typical IR variability of Sgr A*, a detection of a time lag between 0 and -20 minutes arose by chance 2.3\% of the time. Increasing the window of coincidence to include positive time lags (-20 minutes to +20 minutes) resulted in 279 instances of coincidence, or a 4.7\% chance occurrence rate. Figure \ref{fig:hist} shows a the distribution of time lags measured in all \falsePosIterationsTotal simulations. Since we detected a time lag of negative 10--20 minutes in all three of the real \textit{Spitzer}/\textit{Chandra} epochs containing significant X-ray flares, we consider this strong evidence that the X-ray and IR flares from Sgr A* are indeed physically connected. 

\subsection{Comparisons to Previous Work and Flaring Models}

\par
\begin{figure*}
\plotone{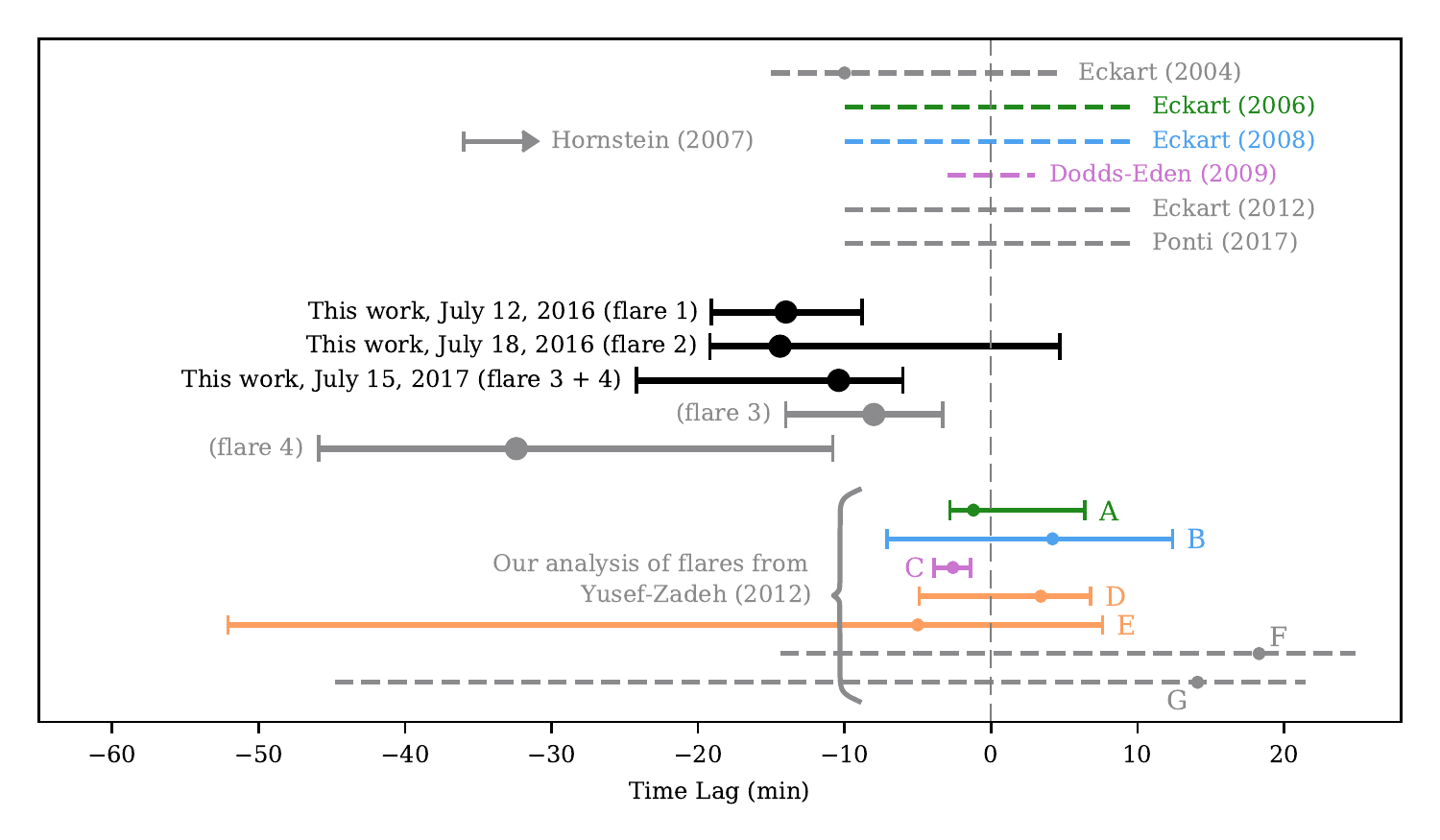}
\caption{Time lags between IR and X-ray flares as reported in this work and in the literature. Plotted in black are the time lags from the three epochs in this work with significant X-ray and IR activity and their 68\% confidence intervals determined from the distribution of \iterations time lags measured from our MC realizations of our light curves. Plotted in solid-grey are the results of the cross-correlation of the isolated sections of the July 15, 2017 light curve containing IR flares 3 and 4. Regions marked with dashed lines come from works that describe the flares to be ``simultaneous to within $x$ minutes" but quote no uncertainties \citep{2004A&A...427....1E,2006A&A...450..535E,2007ApJ...667..900H,2008A&A...479..625E,2009ApJ...698..676D,2012A&A...537A..52E,2017MNRAS.468.2447P}. For example, \cite{2004A&A...427....1E} report an X-ray and IR flare that are simultaneous to within 15 minutes, so we mark that with a line symmetric around zero ranging from $-15$ minutes to $15$ minutes. Several other works report simultaneity between the X-ray and IR peaks, but do not report a time frame within which that claim can be considered valid \citep{2006ApJ...644..198Y,2009ApJ...706..348Y,2011A&A...528A.140T}. The upper limit from \cite{2007ApJ...667..900H} indicates an X-ray flare whose peak occurred 36 minutes before IR observations began. \cite{2012AJ....144....1Y} is the only work to report any time lag between the X-ray and IR with error bars. We re-analyze the seven flares presented in their work and plot the results of our re-analysis here. Five of these flares come from previously reported data sets (color coded as green, blue, magenta and orange for \cite{2006A&A...450..535E}, \cite{2008A&A...479..625E}, \cite{2009ApJ...698..676D}, and \cite{2009ApJ...706..348Y} respectively) and two come from a previously un-reported data set (plotted in grey). The significance of the X-ray flares in these last two data sets is very low (see Section \ref{Sec:re-analysis}).}
\label{fig:timeLag}
\end{figure*}

 There are many models in the literature that attempt to connect Sgr A*'s flares across multiple wavelengths, and particular attention has been paid to the near-simultaneous X-ray and IR peaks. Most models assume the IR peaks are caused by a population of non-thermal electrons emitting synchrotron radiation. Models then evoke synchrotron and/or inverse Compton radiation to connect these IR peaks to their corresponding X-ray flares. Figure \ref{fig:timeLag} displays the time lags found by the cross-correlation analysis in this work in the context of lags (or lack thereof) reported in the literature with flares in both X-ray and IR. Most previous observations with overlapping X-ray and IR coverage reported no time lag between flares seen in both wavelengths, using language like ``simultaneous to within $x$ minutes" but quoting no uncertainties. 

\subsubsection{Synchrotron}

\par In models where both the IR and X-ray emission are produced through direct synchrotron radiation, the population of electrons responsible for the emission would have very high energies ($\gamma \gtrsim10^{5}$), and the cooling timescale would be much shorter than the duration of a typical X-ray flare. This requires that the energized electrons have a sustained source over the duration of the flare \citep{2001Natur.413...45B, 2001A&A...379L..13M, 2004ApJ...606..894Y,2017MNRAS.468.2447P}. Such processes often predict a simultaneous rise in the X-ray and IR flares, with the higher energy X-rays fading faster as the electrons lose energy through synchrotron cooling and/or adiabatic expansion \citep[e.g.,][]{2010ApJ...725..450D}. These predictions from pure synchrotron models are consistent with our tentative measurement of a time lag between the peak in the X-rays and a peak in the IR (equivalently, the X-ray flare rising with the IR but falling faster). Alternatively, if the IR emission is characterized by a number of ``subpeaks" then the X-ray flare could be coincident with an individual subpeak. This could lead to an offset between the X-ray peak and the centroid of the broader IR envelope. 

\par Two studies with observations of simultaneous X-ray and IR flares \citep{2009ApJ...698..676D,2017MNRAS.468.2447P} found the X-ray flares to be consistent with synchrotron radiation. They reported particularly bright X-ray flares coincident with IR peaks where the IR rise had already begun when the X-ray flux rose, and the X-ray flare fell before the IR peak ended. Both of these studies used measurements of the spectral indices of the flares to argue that a synchrotron emission mechanism with a cooling break was responsible for both the IR and X-ray flares. The simultaneous X-ray and IR peaks of \cite{2009ApJ...698..676D} were again interpreted by \cite{2010ApJ...725..450D} in the context of synchrotron emission due to accelerated electrons from magnetic reconnection, and \cite{2017MNRAS.468.2552L} re-interpreted the data in the context of magnetic reconnection accelerating electrons in the coronal region rather than the main body of the accretion flow. 
\par During a low flux density phase of activity, \cite{2018ApJ...863...15W} measured the spectral index of the \textit{Spitzer} IR measurements to be significantly redder than typically observed at high flux densities. Using this measurement as constraint, they determined that the majority of NIR variability data is consistent with a variable spectral index that linearly depends on the logarithm of flux density. Both this determination of the variable IR spectral index for a large sample of variability data and the individual multi-wavelength flare analyses in \cite{2010ApJ...725..450D} and \cite{2017MNRAS.468.2447P} suggest synchrotron mechanisms, which also dominate the submm to IR, and at times reach energies high enough to cause X-ray flares. Indeed, the unchanging spectral index of X-ray variability at high energies is also consistent with a synchrotron scenario \citep{2017ApJ...843...96Z}.

\subsubsection{Synchrotron Self-Compton (SSC)}

In the scenario where the electrons responsible for IR synchrotron radiation are accelerated to energies of $\gamma \sim 100 - 1000$, they could scatter the synchrotron IR photons up to X-ray energies through synchrotron self-Compton (SSC), predicting near simultaneity for the flares. Electrons of these energies in a magnetic field with a strength typically assumed for the accretion flow around Sgr A* would have synchrotron cooling timescales on the order of hours, which fits the duration of the strongest X-ray flares observed in other works \citep[e.g.,][Haggard et al. 2018 (in prep)]{2008A&A...479..625E, 2009ApJ...698..676D, 2017MNRAS.468.2447P}. Several authors have modelled the flares as a SSC process \citep[e.g.,][]{2001A&A...379L..13M,2004A&A...427....1E,2006A&A...450..535E,2008A&A...479..625E,2011A&A...528A.140T,2014MNRAS.441.1005D}.

\par Two of these studies \citep{2006A&A...450..535E,2008A&A...479..625E} reported an X-ray and IR flare to be simultaneous to within 10 minutes. \cite{2006A&A...450..535E} used this simultaneity and the suggestion that the IR flare spectrum is relatively red with a variable spectral index \citep[e.g.,][]{2005ApJ...628..246E,2018ApJ...863...15W} to argue in favour of the SSC picture. \cite{2008A&A...479..625E} observed a polarized IR flare with an X-ray counterpart. They also found that the flares fit the SSC picture of submm synchrotron photons being up-scattered to IR and X-ray wavelengths and discussed this in the context of a model involving a temporary disk and a short jet, where rotating spots in the disk  are responsible for variations in the IR flares.

\cite{2004A&A...427....1E} reported a possible time lag of $\sim$10 minutes between the first flares detected in simultaneous observations, with the X-ray leading the IR. Due to the large binning of their X-ray light curves they quote $\sim$15 minutes as a conservative upper limit for any time lag. They describe the flares with a SSC model in which IR flares are due to both synchrotron and the up-scattering of synchrotron submm photons, and X-ray flares are produced through the IC scattering of  submm and IR photons.

Finally, reporting no constraints on the simultaneity of the X-ray and IR variability in their work, \cite{2011A&A...528A.140T} interpreted their X-ray/IR/submm data in the context of a model involving an expanding plasmoid releasing synchrotron submm and IR radiation and up-scattering to X-ray energies through SSC processes. These models are often invoked to explain the tentative time lags between the X-ray/IR flares and submm or radio activity. The authors found that the simplest version of this expanding plasmoid model did not adequately explain their data. 

\subsubsection{Inverse Compton (IC)}

In IC models connecting the IR and X-ray variability, IR-synchrotron emitting electrons with energies $\gamma \sim 100 - 1000$ scatter submm seed photons up to X-ray energies, predicting a potential lag in the X-rays relative to the IR outburst. This is inconsistent with our measurement of a negative X-ray time lag, which illustrates (at least tentatively) that the X-rays \textit{lead} the IR, or at the very least, rise simultaneously and fall faster. 

\begin{figure*}[ht]
  \centering
  \includegraphics[width=0.95\textwidth]{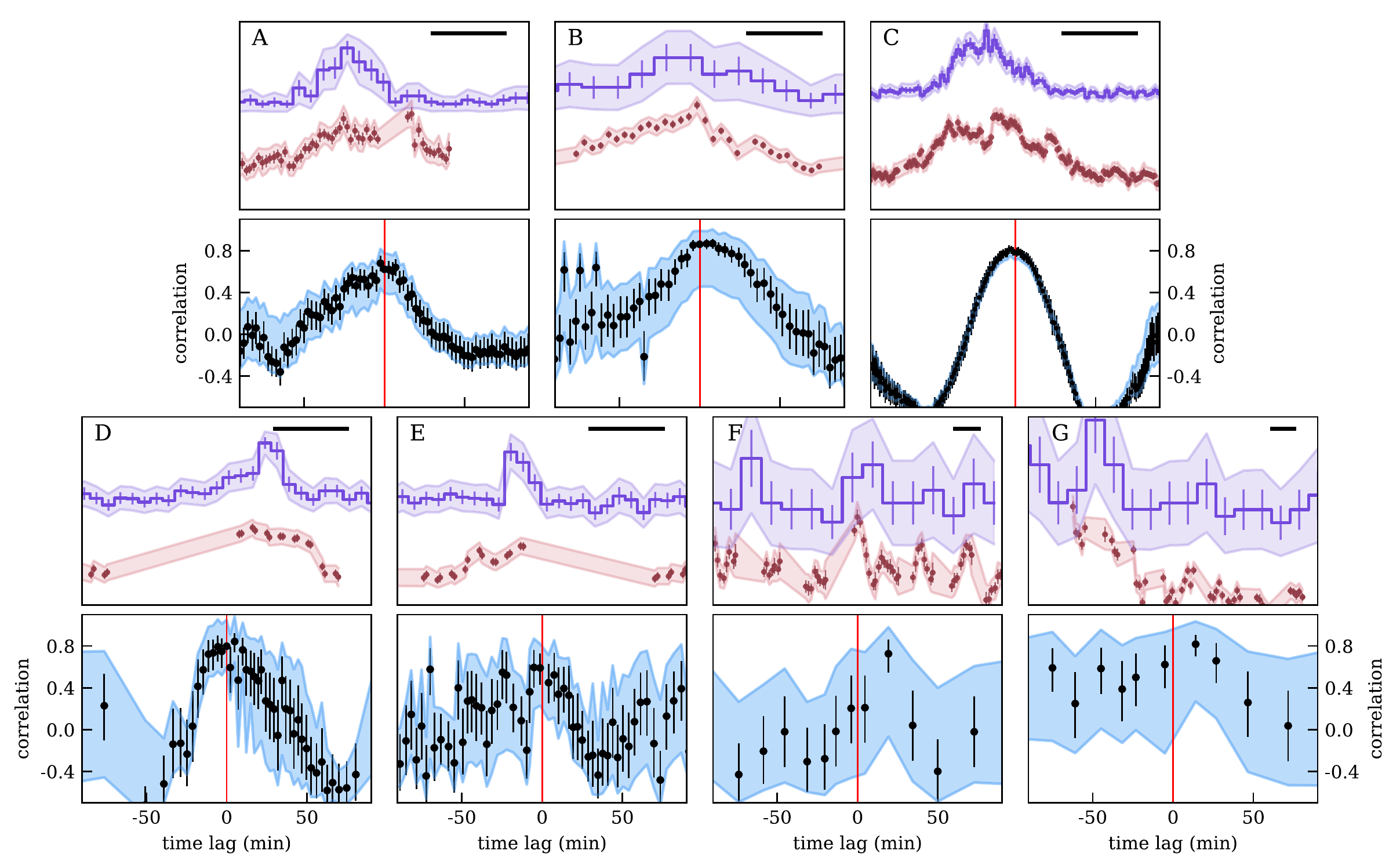}
  \caption{X-ray and IR light curves discussed in \cite{2012AJ....144....1Y} and the cross-correlations we find in our MC analysis. First and third row panels display the light curves along with the 95\% envelopes of our MC realizations generated by drawing from a normal distribution centered around the true data with a standard deviation proportional to the errors on the light curves. Symbols and line styles are identical to those in Figure \ref{fig:zdcf}. For this visualization, the y-scales are arbitrary, and the black line in the upper right of each white panel indicates a 30 minute interval. The results of the \texttt{ZDCF} on the real light curves are displayed in black in the corresponding lower panels, while the MC envelope from which the time lags are measured is displayed in blue. The resulting time lags and confidence intervals are reported in column five of Table \ref{tab:lit_timeLags} and plotted in Figure \ref{fig:timeLag}. Table \ref{tab:lit_timeLags} lists the dates, original papers, and facilities from which the IR and X-ray light curves come. All X-ray light curves were binned at 300 s except for F and G, which were binned at 1500 s. IR light curves A, B, and C were binned at 140 s, while light curves D and E were binned at 144 s and light curves F and G were binned at 120 s. \label{fig:YZ+2012_allFlares}}
\end{figure*}

\par \cite{2006ApJ...644..198Y} and \cite{2009ApJ...706..348Y} both detected several faint X-ray flares with IR counterparts, but quoted no uncertainty on any time lag.
 \cite{2006ApJ...644..198Y} predicted that in the IC scenario, the spectral index of an IR flare must be the same as the correlated X-ray flare (i.e., $\alpha$$\sim$0.6). Unfortunately, the X-ray flare they observed did not contain enough counts to provide any spectral information. In their campaign to observe Sgr A* across many wavelengths (cm, mm, submm, IR, X-ray), \cite{2009ApJ...706..348Y} also argued in favour of an IC model, but one in which synchrotron IR photons are up-scattered to X-ray wavelengths by electrons responsible for the radio and submm emission of Sgr A*. 
 \par As the only other work to quote uncertainties on the time lags found in their cross-correlations, \cite{2012AJ....144....1Y} re-analyzed archival data \citep{2006ApJ...644..198Y, 2006A&A...450..535E, 2008A&A...479..625E, 2009ApJ...706..348Y, 2009ApJ...698..676D} along with previously un-reported observations. They reported evidence for the peak of the X-ray flares \textit{lagging} the IR peaks with a time delay ranging from a few to tens of minutes. Assuming there is not more than one population of flares, this is in tension with our finding of the X-rays tentatively \textit{leading} the IR. These authors also employed the \texttt{ZDCF}, and the time lags they find are quoted in Table \ref{tab:lit_timeLags}.

\begin{table*}[t]

	\begin{center}
	\caption{Time Lags: \cite{2012AJ....144....1Y} Flares }
	\label{tab:lit_timeLags}
	\begin{tabular}{cllllrr}
	
	\toprule
	Flare & Date  &  Orig. Paper  & IR Facility & X-ray Facility & \multicolumn{1}{c}{\begin{tabular}[c]{@{}c@{}}YZ\\ time lag\\ (min)\end{tabular}} &  \multicolumn{1}{c}{\begin{tabular}[c]{@{}c@{}}Our\\ time lag\\ (min)\end{tabular}} \\
	\midrule
	A & 2004 Jul 6/7 & Eck+2006b & \textit{VLT} (K band) & \textit{Chandra} & $7\substack{+1.3 \\ -1.2}$ & $-1.2\substack{+7.6 \\ -1.6}$ \\
	B & 2005 Jul 30 & Eck+2008 & \textit{VLT} (K band) & \textit{Chandra} & $8\substack{+10 \\ -10.1}$ & $4.2\substack{+8.2 \\ -11.3}$ \\
	C & 2007 Apr 4 & DE+2009 & \textit{VLT} (K band) & \textit{Chandra} & $-0.5\substack{+7 \\ -6.5}$ & $-2.6\substack{+1.2 \\ -1.3}$ \\
	D & 2007 Apr 4 & YZ+2009 & \textit{HST}/NICMOS & \textit{XXM-Newton} & $5\substack{+1 \\ -1.4}$ & $3.4\substack{+3.4 \\ -8.3}$ \\
	E & 2007 Apr 4 & YZ+2009 & \textit{HST}/NICMOS & \textit{XXM-Newton} & $5.0\substack{+1.9 \\ -1.5}$ & $-5.0\substack{+12.6 \\ -47.1}$ \\
	F & 2008 May 5 & YZ+2012 & \textit{VLT} (K band) & \textit{Chandra} & $19\substack{+6.8 \\ -2.4}$ & $18.3\substack{+6.6 \\ -32.7}$ \\
	G & 2008 Jul 26/27 & YZ+2012 & \textit{VLT} (K band) & \textit{Chandra} & $14.6\substack{+5.6 \\ -7.4}$ & $14.1\substack{+7.4 \\ -58.9}$ \\
	\bottomrule
	\end{tabular}
	\end{center}
	
\justifying
\textit{Note}: Negative values mean X-ray leads IR. The first column labels the flares. The second column lists the date the simultaneous data was taken. Column three lists the original paper that the data was reported in (Eck+2006b = \cite{2006A&A...450..535E}, Eck+2008 = \cite{2008A&A...479..625E}, DE+2009 = \cite{2009ApJ...698..676D}, YZ+2009 = \cite{2009ApJ...706..348Y}, YZ+2012 = \cite{2012AJ....144....1Y} ). Column four and five list the facilities with which the IR and X-ray data were collected. Column six lists the time lag and 1$\sigma$ errors reported from the \texttt{ZDCF} analysis in \cite{2012AJ....144....1Y}. The last column lists the time lags and 68\% confidence intervals we find from our Monte Carlo analysis (\lititerations realizations).
\end{table*}

\vspace{5mm}
\subsection{Re-analyzing Light Curves from the Literature}\label{Sec:re-analysis}
Due to the very low signal-to-noise in the individual cross-correlation results and the large binning of some of the X-ray data in \cite{2012AJ....144....1Y}, we suspect their reported 1$\sigma$ error bars to be underestimated. For these reasons, we elected to run their light curves through our cross-correlation and Monte Carlo analysis to verify their results, provide us with a consistent comparison, and determine the effect that the signal-to-noise ratio of a flare has on the cross-correlation. 
\par Using the seven light curves presented in \cite{2012AJ....144....1Y}, we performed a cross-correlation analysis identical to the analysis we applied to our own light curves in section \ref{CrossCorrelation}. We generated \lititerations Monte Carlo instances of the light curves scaled in proportion to the errors on the data and ran the \texttt{ZDCF} on them. The mean of the resulting distribution of time lags was adopted as our measurement, with uncertainties determined by the interval within which 68\% of the time lags fall. Table \ref{tab:lit_timeLags} labels the flares A to G and displays the time lags and 1$\sigma$ errors reported by \cite{2012AJ....144....1Y} as well as the time lags and uncertainties that we measure from our MC analysis. Figure \ref{fig:YZ+2012_allFlares} re-plots the light curves found in their work (with the exception of Flare C, which is not plotted by \cite{2012AJ....144....1Y} but still included in their analysis) along with the results of our Monte Carlo analysis with the \texttt{ZDCF}. Figure \ref{fig:timeLag} displays our measurements for these seven flares in the context of this work and the literature.
\par The differences between our results and those reported by \cite{2012AJ....144....1Y} are due to our MC analysis incorporating the signal-to-noise of the input light curves. In a light curve with low signal-to-noise, larger uncertainties on the data points will produce MC simulations that span a larger flux range, resulting in a broader range of time lags found between features in the light curves. Both methods employ the \texttt{ZDCF}, but while \cite{2012AJ....144....1Y} quote the time lag of a single \texttt{ZDCF}/\texttt{PLIKE} run, we measure the time lag and estimate the uncertainties from \iterations runs of the \texttt{ZDCF}.
\par In comparison to \cite{2012AJ....144....1Y}, this method estimates larger and more realistic uncertainties on the time lags for flares A, D, E, F, and G, similar uncertainties for flare B, and a smaller uncertainty for the time lag of Flare C, the brightest simultaneous X-ray and IR flare observed to date \citep{2009ApJ...698..676D}. Our analysis of flares A, C, D, and E found the time lag to be closer to zero or even negative compared to the lag reported by \cite{2012AJ....144....1Y}. The low signal-to-noise in the light curves of Flares E, F and G is reflected in the poor constraints on a time-lag. 

\par
\begin{figure}[ht]
  \centering
  \includegraphics[width=0.48\textwidth]{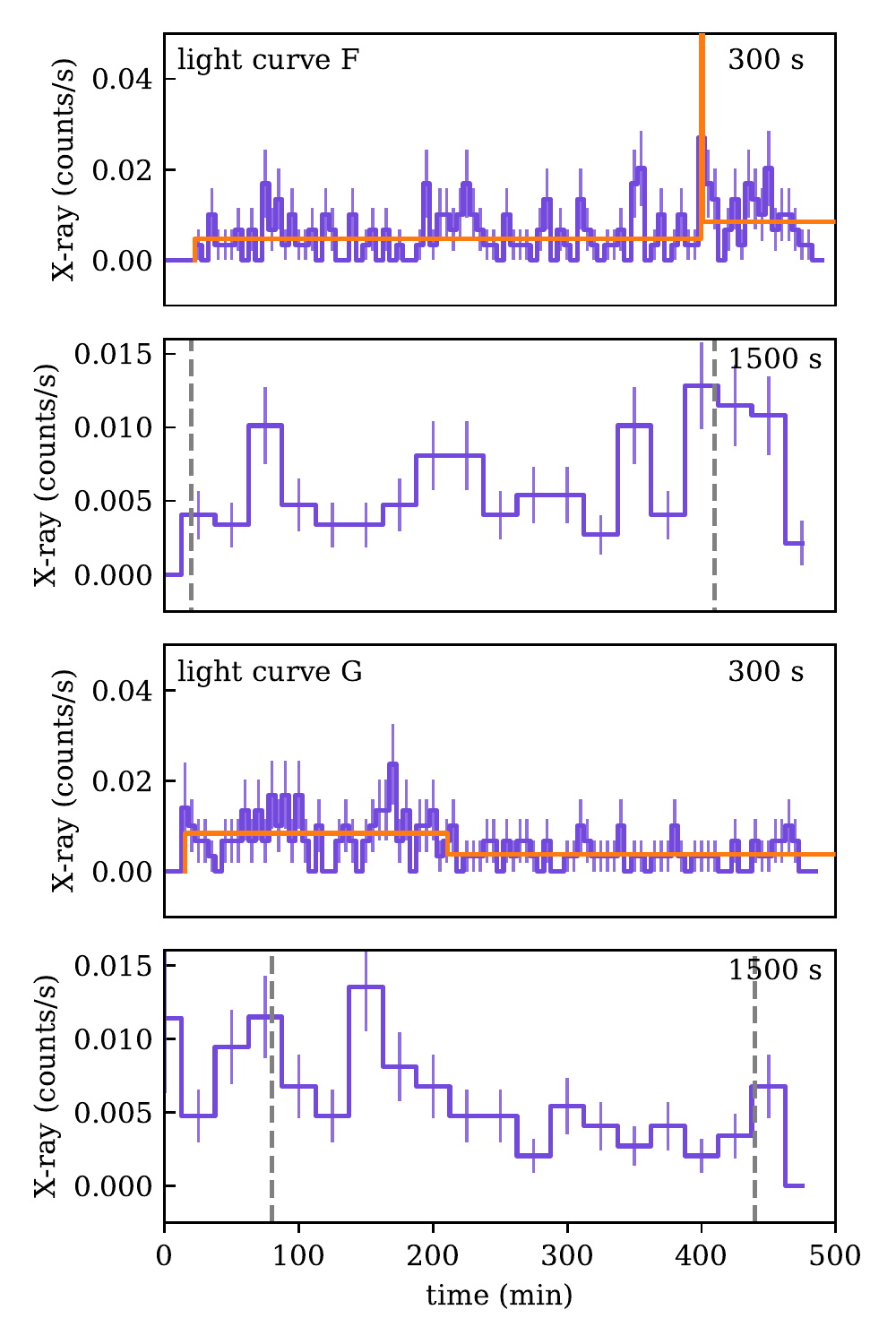}
  \caption{Sgr A* X-ray light curves extracted from \textit{Chandra} ObsID 9169 (light curve F) and ObsID 9173 (light curve G). \textit{Top}: Light curves with 300 s binning. Orange displays the Bayesian Blocks results with p$_{0}=$0.1 \textit{Bottom}: Light curves with 1500 s (25 min) binning. Grey dashed lines indicate the same interval analyzed by \cite{2012AJ....144....1Y}. \label{fig:flareFandG}}
\end{figure}

 In the case of flares F and G, we question whether cross-correlating these X-ray light curves binned at 25 minutes with IR light curves binned at 2 minutes is meaningful. To test the variability of the light curves containing flares F and G, we opted to download the raw \textit{Chandra} data and run the flare detection algorithm we used for our X-ray light curves (Bayesian Blocks algorithm, see Section \ref{FlareAnalysis}). The data are found within \textit{Chandra} ObsIDs 9169 and 9173. Figure \ref{fig:flareFandG} plots the results of running the Bayesian blocks flare detection algorithm on light curves F and G, the lowest signal-to-noise light curves in Figure \ref{fig:YZ+2012_allFlares}.  In our Bayesian blocks analysis of these light curves (orange lines in Figure A2), we do not detect any statistically significant X-ray peaks near the IR flares discussed by \cite{2012AJ....144....1Y}. The cross-correlations for these observations should therefore not be considered measurements of meaningful lags between IR and X-ray variability.

\vspace{5mm}
\section{Conclusion}

\par We have presented results based on new observations which are the longest simultaneous IR and X-ray observations of Sgr A* to date. These overlapping light curves offer the best tests of the connection between these two wavelengths, and provide a crucial probe of Sgr A*'s variable emission. We detect four X-ray flares ($\sim$4 $\times$ quiescence) and no ``strong" X-ray flares ($\ge$10 $\times$ quiescence) during the combined $>$100 hours that Chandra observed Sgr A*. The IR emission of Sgr A* showed peaks coinciding with the weak X-ray flares and also occurring at times when no X-ray flares are detectable. A cross-correlation analysis of all our simultaneous light curves suggests that the X-ray flares may \textit{lead} the IR by approximately 10--20 minutes, but the 99.7\% confidence intervals are still consistent with zero time-lag. This is in agreement with models that describe both the X-ray and IR flares as synchrotron emission originating from particle acceleration events involving magnetic reconnection and shocks in the accretion flow \citep[e.g., see 4.1 of ][]{2010ApJ...725..450D} and consistent with models that predict simultaneity of the flares through SSC processes. Our results are inconsistent with models invoking external populations of electrons through IC processes as described by \cite{2012AJ....144....1Y}, though it is not obvious that all X-ray/IR flares are produced by the same process. 

\par It remains difficult to distinguish between the suggested flaring mechanisms connecting the X-ray and IR. Despite having the longest uninterrupted and simultaneous X-ray/IR dataset of Sgr A* to date, we observed no bright X-ray flares during the 4+ days of observations reported here. Though this prevented us from gaining spectral information from the faint X-ray flares, future coordinated observations may catch significantly brighter simultaneous outbursts as has happened in the past \citep{2009ApJ...698..676D} and will certainly add to the growing statistical strength of the catalog of multi-wavelength flares. Previous observations have revealed substructure in the outbursts of both wavelengths \citep[e.g.,][Haggard et al. 2018 (in prep)]{2009ApJ...698..676D}, and future observations of bright simultaneous outbursts could allow for a more detailed cross-correlation of sub-components in the X-ray and IR peaks.

\par In the immediate future, upcoming \textit{Spitzer}/\textit{Chandra} observations approved for the Summer of 2019\footnote{\href{https://www.cfa.harvard.edu/irac/gc/}{https://www.cfa.harvard.edu/irac/gc/}} may detect multiple bright flares, which may be key to constraining the time lag between the X-ray and the IR. In the longer term, a better understanding of the time-dependent emission from Sgr A* will allow for the characterization of the accretion physics around the black hole and inform the next generation of GRMHD simulations. Not only will long epochs of observations at multiple wavelengths be ideal data sets for distinguishing between semi-analytical flaring models, but Sgr A*'s variability will provide a strict benchmark for testing whether or not state-of-the-art simulations are probing the real physical scales of the turbulent accretion flow.  Additionally, the multi-wavelength efforts coordinated with the Event Horizon Campaigns in April 2017 and 2018 hold promise for narrowing in on the physical processes that drive Sgr A*'s variability. 

\acknowledgments

The scientific results reported in this article are based on observations made by the Chandra X-ray Observatory and the Spitzer Space Telescope. We thank the Chandra and Spitzer scheduling, data processing, and archive teams for making these observations possible. Support for this work was provided by the National Aeronautics and Space Administration through Chandra Award Number GO7-18135B issued by the Chandra X-ray Center, which is operated by the Smithsonian Astrophysical Observatory for and on behalf of the National Aeronautics Space Administration under contract NAS8-03060. D.H. acknowledges support from a Natural Sciences and Engineering Research Council of Canada (NSERC) Discovery Grant and a Fonds de recherche du Qu\'{e}bec--Nature et Technologies (FRQNT) Nouveaux Chercheurs Grant. G.W. acknowledges support from the NSF grants AST-0909218, AST-1412615. JLH acknowledges support from NASA Grant 80NSSC18K0416. GP acknowledges financial support from BMWi/DLR grants FKZ 50 OR 1604, 50 OR 1715 and 50 OR 1812.\\
\textit{Software}: CIAO \cite{2006SPIE.6270E..1VF}, NumPy \citep{SciPy}, AstroPy \citep{2018arXiv180102634T}, Matplotlib \citep{Hunter:2007}, Bayesian Blocks \citep{2017ascl.soft04001W}, ZDCF \citep{2013arXiv1302.1508A} \\
\textit{Facilities}: Spitzer/IRAC, Chandra/ACIS

\bibliographystyle{aasjournal}
\bibliography{boyce_SgrA_IR-Xray}{}

\end{document}